\documentclass[12pt, draftclsnofoot, onecolumn]{IEEEtran}
\usepackage{setspace}
\usepackage{cite}

\usepackage{graphicx}
\graphicspath{/figs/}
\DeclareGraphicsExtensions{.pdf,.jpeg,.png}
\usepackage{amsmath, amssymb}

\usepackage[table]{xcolor} % can add monochrome
\usepackage{bm}

\usepackage{float}
\usepackage[caption = false]{subfig}
\usepackage{algorithm}
\usepackage{algorithmicx}
\usepackage{algpseudocode}
\usepackage{pifont}
\usepackage{xcolor}
\usepackage{relsize}

\usepackage{array}
\usepackage{url}
\usepackage{parskip}
\usepackage{layout}

\let\OLDthebibliography\thebibliography
\renewcommand\thebibliography[1]{
  \OLDthebibliography{#1}
  \setlength{\parskip}{0pt}
  \setlength{\itemsep}{5pt plus 0.3ex}
}

\usepackage{mathtools}
\DeclarePairedDelimiter{\ceil}{\lceil}{\rceil}
\DeclarePairedDelimiter{\floor}{\lfloor}{\rfloor}
\newcommand{\round}[1]{\ensuremath{\lfloor#1\rceil}}
\DeclareMathOperator*{\argmax}{arg\,max}

\algdef{SE}[DOWHILE]{Do}{doWhile}{\algorithmicdo}[1]{\algorithmicwhile\ #1}%

\newcommand{\ue}{u \color{black}}

\newcommand{\analog}[1]{#1^{\text{RF}}}
\newcommand{\Nxa}{\analog{N}_{\text{X}}}
\newcommand{\Nya}{\analog{N}_{\text{Y}}}

\newcommand{\Ntd}{N_{\text{Td}}}

\newcommand{\Nx}{N_{\text{X}}}
\newcommand{\Ny}{N_{\text{Y}}}
\newcommand{\Nt}{N_{\text{T}}}
\newcommand{\Nta}{\analog{N}_{\text{T}}}
\newcommand{\Nr}{N_{\text{R}}}

\newcommand{\nt}{n_{\text{t}}}
\newcommand{\nr}{n_{\text{r}}}
\newcommand{\Np}{N_{\text{P}}}
\newcommand{\Oh}{O_{\text{H}}}
\newcommand{\Ov}{O_{\text{V}}}

\newcommand{\norm}[1]{\left\lVert#1\right\rVert}

\newcommand{\ji}{\mathsf{j}}
\newcommand{\vander}[2]{\mathbf{a}_{N_{#2}}(#1)}
\newcommand{\vect}[1]{\mathbf{#1}}
\newcommand{\complex}{\mathbb{C}}

\newcommand{\phaseres}{b_{\text{phase}}}
\newcommand{\W}{\mathbf{W}}
\newcommand{\Wutk}{\mathbf{W}_{\ue, t, k}}
\newcommand{\F}{\mathbf{F}}
\newcommand{\Futk}{\mathbf{F}_{\ue, t, k}}
\newcommand{\f}{\mathbf{f}}

\newcommand{\Fa}{\analog{\mathbf{F}}_{t}}

\newcommand{\Fistar}{\F_{\vect{\hat{i}}_u}}
\newcommand{\Fssb}{\mathbf{F}^{\text{SSB}}}

\newcommand{\Bssb}{\mathbf{B}^{\text{SSB}}}
\newcommand{\Bcsi}{\mathbf{B}^{\text{CSI-RS}}}

\newcommand{\Bactive}{\mathbf{B}^{\text{sub}}}
\newcommand{\Lmax}{L_{\text{{max}}}}
\newcommand{\Kssb}{K^{\text{SSB}}}
\newcommand{\Tssb}{T^{\text{SSB}}_{i}}

\newcommand{\ycsirs}{\vect{y}^{\text{CSI-RS}_i}}
\newcommand{\ycsirshat}{\vect{y}^{\text{CSI-RS}_{\hat{\vect{i}}_u}}}
\newcommand{\Kcsi}{K^{\text{CSI-RS}_i}}
\newcommand{\Tcsi}{T^{\text{CSI-RS}_{i}}}
\newcommand{\Lcsi}{L_{\text{{CSI}}}}
\newcommand{\bwp}{\text{S}_{\text{B}}}
\newcommand{\Pcsi}{P_{\text{CSI}}}
\newcommand{\nRB}{N_{\text{RB}}}
\newcommand{\Ncsi}{N_{\text{CSI}}}
\newcommand{\Bg}{B_{\text{g}}}
 \newcommand{\Obsc}{\vect{O}^{\text{BSC}}}
\newcommand{\Achan}{\vect{H}}

\newcommand{\algo}{\text{X-BM}}
\newcommand{\tku}[1]{{#1_{\ue, t, k}}}

\newcommand{\RSRPiu}{\text{RSRP}_{i, u}}
\newcommand{\SNR}{\text{SNR}}
\newcommand{\SINR}{\text{SINR}}

\begin{document}

\title{Hierarchical ML Codebook Design for Extreme MIMO Beam Management}

\author{\IEEEauthorblockN{Ryan M. Dreifuerst,~\IEEEmembership{Graduate Student Member,~IEEE, and }}%
    \and
	\IEEEauthorblockN{Robert~W.~Heath~Jr.~\IEEEmembership{Fellow,~IEEE}}%

	\thanks{Ryan M. Dreifuerst and Robert W. Heath Jr. are with North Carolina State University, Raleigh, NC 27695 \{rmdreifu, rwheathjr\}@ncsu.edu.
	       This material is based upon work supported by the National Science Foundation under grant nos. NSF-ECCS-2153698, NSF-CCF-2225555, NSF-CNS-2147955  and is supported in part by funds from federal agency and industry partners as specified in the Resilient \& Intelligent NextG Systems (RINGS) program.}% <-this % stops a space
}

\maketitle
\bstctlcite{IEEEexample:BSTcontrol}

\begin{abstract}
    Beam management is a strategy to unify beamforming and channel state information (CSI) acquisition with large antenna arrays in 5G. Codebooks serve multiple uses in beam management including beamforming reference signals, CSI reporting, and analog beam training. 
    % The beam management framework follows the same strategy as used in millimeter-wave 5G which includes initial access, refinement, and data transmission stages. 
    In this paper, we propose and evaluate a machine learning-refined codebook design process for extremely large multiple-input multiple-output (X-MIMO) systems. We propose a neural network and beam selection strategy to design the initial access and refinement codebooks using end-to-end learning from beamspace representations. The algorithm, called Extreme-Beam Management  ($\algo$), can significantly improve the performance of extremely large arrays as envisioned for $6$G and capture realistic wireless and physical layer aspects. Our results show an $8$dB improvement in initial access and overall effective spectral efficiency improvements compared to traditional codebook methods.
    
\end{abstract}

\section{Introduction}
    % paragraph 1, massive MIMO in 5G advanced and 6G
    Large-scale or Massive MIMO (M-MIMO) has been an important technology in mobile broadband since the first investigations on $3$D beamforming in 3GPP release $13$ \cite{Dreifuerst2023magazine}. Since then, larger arrays have been envisioned with extremely large arrays ($\gg64$ antennas) for 6G \cite{Holma2021XMIMO}.
    Effectively employing M-MIMO or X-MIMO, however, requires obtaining accurate CSI. 
    With the inception of 5G, M-MIMO was supported through a process called beam management which integrated beamforming for control signaling, hierarchical codebooks for beam training, and enabled multiple configurations of feedback \cite{Dreifuerst2023magazine}. 
    Notably, beam management supports hybrid arrays where the number of antenna ports visible to digital communication systems and the number of antennas in the array might be vastly different, as found in the so-called hybrid MIMO architecture \cite{HeathOverviewSP4mmWave2016}. 
    While the beam management framework is designed to be flexible, there has been limited research on optimizing beam management codebooks for this flexible framework.

    % Beam management
    Beam management is a multi-step process that includes transmitting beamformed reference signals and obtaining multiple forms of CSI to balance the overhead, latency, and accuracy \cite{HengSixChallengesBM6G2021, Giordani3GPPBeamManagement2019, Dreifuerst2023magazine}. In 5G, the first step is synchronization which requires transmitting one or more synchronization signal blocks (SSB) with a beamformer from a codebook (denoted as the \textit{SSB codebook}). These are typically understood as wide-area coverage beams because synchronization does not require a high signal-to-noise ratio (SNR), but wide-area beamforming improves cell coverage and aids the subsequent beam search. From the set of SSBs, user equipment (UE) provides a small feedback message identifying the strongest SSB and the power received from the transmission \cite{Giordani3GPPBeamManagement2019}. Based on the SSB feedback, the base station (BS) selects a refined beamforming codebook (identified as the \textit{CSI-RS codebook}) for transmitting CSI reference signals (CSI-RS), which contain pilot signals for channel estimation. These two beamforming codebooks support different functions (initial access and synchronization compared to beam refinement and channel estimation) and operate over different time-frequency resource allocations to support these roles. Following CSI-RS reception, UEs estimate the effective channel and quantize it according to another codebook (the \textit{feedback codebook}) based on the BS array geometry. Finally, the CSI feedback is provided to the BS and data transmission can begin with an appropriately designed precoder to serve one or more users. 

    % Machine Learning in Wireless
    Optimizing the various codebooks is challenging due to the difficulty in mathematically capturing the relationships between the channel, mobile users, and the codebook interrelationships. One possible solution is applying artificial intelligence or machine learning (AI/ML) to learn these underlying characteristics from data. While AI/ML has become an integral part of domains like computer vision and language modeling, mobile broadband has not seen as many benefits due to the well-known structure and strict timing requirements of wireless communications \cite{WangEtAlDL4PHYSurvey2017, ZhangDLinWirelessSurvey2019}. Recently, 3GPP has announced investigations for the integration of AI/ML targeting 3GPP Release $18$ and beyond \cite{Lin2023OverviewAI5G}. In particular, beam management and CSI feedback are identified as key areas where learning strategies are envisioned \cite{Lin2023OverviewAI5G}. 

    % 6G
    With the initial proposals for 6G underway, new research on AI/ML for extremely large arrays, equipped with hundreds of antennas, is a critical topic, especially operating in new communication bands \cite{Holma2021XMIMO}. 5G initially began supporting massive MIMO in a hybrid format, where analog beam training was performed using the SSB and CSI-RS beamformers, and digital precoding was based on the CSI feedback. Given the support and success of these formats, it is valuable to design and optimize strategies for X-MIMO within this formulation. Successfully accomplishing this task paves the way for beam management with AI/ML in 6G.
    % we assume a similar strategy, although we will carefully consider the efficiency and efficacy when scaled up to X-MIMO systems. 

    % References for beam management or beam training?
    There are three directions of work related to AI/ML beam management: 1) beam training, 2) codebook design, and 3) CSI feedback. First, we review a set of relevant prior work incorporating machine learning and beam training. Beam training has received the most attention of the three directions when it comes to machine learning applications. Classical data-driven techniques were explored in \cite{LiBeamTraining2013, V.VaEtAlInverseMultipathFingerprintingMillimeter2018} using historical beam training data. While the codebooks and algorithms rely on more traditional techniques, the concept of ``learning'' a reduced codebook from site-specific data is a foundational concept for more recent deep learning techniques \cite{WangEtAlSiteSpecificCompressiveCodebook2021, Xue2022BMFRL, Yang2022DLBeamAlignment, Yang2023hierarchicalBeamAlignment}. These papers (\cite{WangEtAlSiteSpecificCompressiveCodebook2021, Xue2022BMFRL, Yang2022DLBeamAlignment, Yang2023hierarchicalBeamAlignment}) focus on millimeter-wave (mmWave) channels and typically exploit the sparsity to aid the beam training process. Furthermore, the codebooks within these papers are standard codebooks like DFT which are unlikely to perform as well in rich scattering environments as seen in upper-mid bands ($7$-$20$GHz) envisioned for 6G \cite{Holma2021XMIMO}. In addition, that work \cite{WangEtAlSiteSpecificCompressiveCodebook2021, Xue2022BMFRL, Yang2022DLBeamAlignment, Yang2023hierarchicalBeamAlignment} does not consider a MIMO format with multiple UE antennas and streams of data, instead limiting the setting to analog beamforming. In this paper, we consider a scenario with multiple users, multiple data streams, and frequency selective, OFDM channels in rich multipath environments.

    % Reference for codebook design
    Codebook and beamformer design using machine learning has received some focus in recent years \cite{ShafinEtAlDRL4BeamOptim2020, XiaMISOBFwithDL2020, Xue2022BMFRL, HengProbingBeams2021, Dreifuerst2022VTCBM}. Hierarchical codebooks, which are a core concept in 5G, were proposed in \cite{OrigHierarchical2009, Alkhateeb2014, XiaoCodebook2016} to separate the beam search process into multiple resolutions to reduce the overhead of beam search. Deep reinforcement learning for broadcast beam pattern transmission was shown to be effective in small-scale MIMO \cite{ShafinEtAlDRL4BeamOptim2020}. Supervised learning for SSB codebook design in narrowband MIMO was also shown to improve over standard codebooks \cite{Dreifuerst2022VTCBM}. Unsupervised learning was applied to beamformer design from narrowband channel measurements \cite{XiaMISOBFwithDL2020} but required perfect channel state information. Broadcast probing beam patterns were learned using channel state information in \cite{HengProbingBeams2021} which is one of the only papers to consider a hierarchical strategy similar to 5G based on SSB and CSI-RS codebooks. This work \cite{HengProbingBeams2021} provides motivation for researching realistic deep learning codebook methods, but it remains unclear how the integration of ML impacts the overall system performance as opposed to simply improving the beam alignment. Our work addresses this question while further increasing the role that data-driven algorithms play in codebook operations.

    % CSI feedback 
    While codebook design is typically focused on improving the received signal power, codebooks are also employed for quantizing channel state information into discrete feedback. ML has also been applied to improve feedback quantization in recent work \cite{WenDL4mMIMOCSIFeedback2018, KimLearningMISOBF2021, Chen2021, Xiao2023MetaCSI}. In general, ML methods propose encoder-decoder structures \cite{WenDL4mMIMOCSIFeedback2018, KimLearningMISOBF2021, Chen2021, Xiao2023MetaCSI}, that employ jointly-trained models at the UE and BS. This type of application involves significant overhead, however, to ensure both terminals have synchronized neural network weights or parameters. Furthermore, the system effects in MU-MIMO settings are not considered, even though high resolution CSI feedback is primarily designed for improving MU-MIMO performance \cite{Morozov2016}. Throughout these investigations, CSI optimization is treated as a separate task beam training and channel estimation, thereby neglecting the relationships throughout the system. In this paper, we incorporate beamformed channel feedback as a result of beam management to study the effects of ML codebooks on the overall system performance.

    \begin{figure}[!t]
	    \centering
	    \includegraphics[width=6.25in]{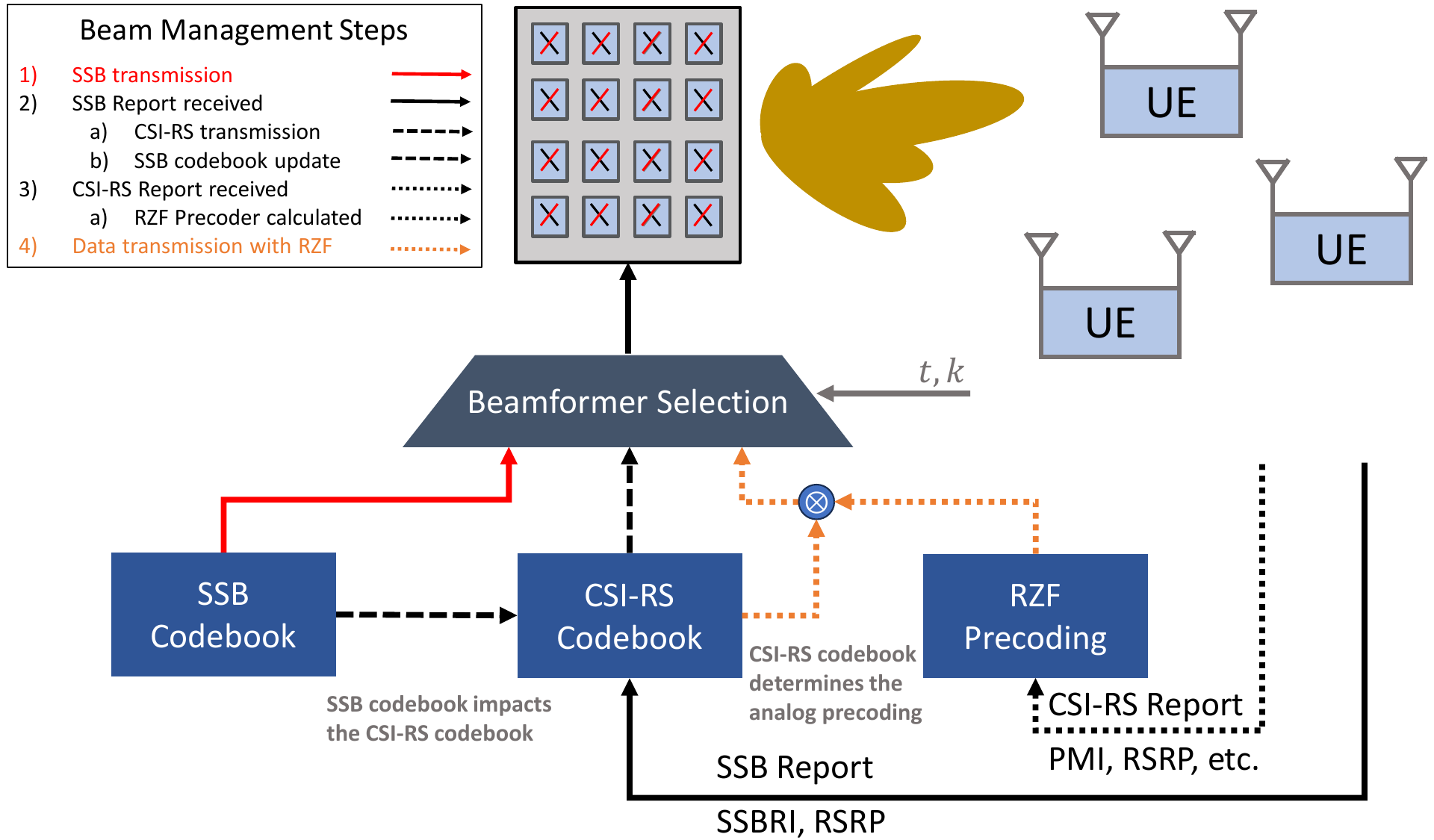}
	    \caption{A depiction of the timing and interaction between the codebooks and data transmission. The SSB and CSI-RS codebooks are used to determine an effective analog precoding that is combined with the precoding determined by the CSI-RS report.}
	    \label{fig: system_model}
	\end{figure}

    In our prior work \cite{Dreifuerst2023CodebookFeedback}, we focused on a 5G, fully digital sub-6GHz system. From that investigation, we found that significant gains can be achieved with improvements to the beam management framework using deep learning. While the strategy was effective for SSB transmission, we now extend the consideration to a new system configuration that supports hybrid, upper-mid band systems with jointly-learned SSB and CSI-RS codebooks. Throughout the process, the overall goal is to balance the computational complexity, cost, and performance of CSI acquisition to maximize system-level performance. Our contributions in this manuscript are as follows:
   \begin{itemize}
        \item First, we precisely define the beam management process for an envisioned hybrid, X-MIMO system. The process includes all relevant steps from initial access up through achievable network spectral efficiency calculation. This system enables a holistic evaluation of codebooks that captures the interdependencies of a realistic beam management strategy for 6G and is a natural evolution of that supported in 5G.
        
        \item Second, we design a multi-task neural network architecture and processing pipeline to design SSB and CSI-RS codebooks. The pipeline starts by converting the current SSB codebook and SSB feedback into the beamspace. Then the input is fed into two neural networks that are jointly trained in an end-to-end fashion to output an SSB and a set of candidate CSI-RS codebooks. This architecture is formulated to fit the strict timing requirements of beam training. The joint neural network formulation is able to learn codebooks that achieve beam training results similar to if perfect CSI were known a priori.
        
        \item Third, we rigorously evaluate the capabilities of the beam management framework in X-MIMO using various codebooks, physically array geometries, and feedback strategies. The most important result we find is that the proposed system, with limited feedback, can achieve comparable performance to DFT-codebook systems with full feedback in MU-MIMO settings. Furthermore, we find that the $\algo$ design transfers efficiently between new environments, antenna arrays, and frequency ranges. 
    \end{itemize}

% Notation
    $\mathbf{Notation}$: $\vect{A}$ is a matrix, $\vect{a}$ and $\{a[i]\}_{i=1}^{N}$ are column vectors and $a,A$ denote scalars. $\vect{A}^T$, $\overline{\vect{A}}$, $\vect{A}^*$, and $\vect{A}^{\dagger}$ represent the transpose, conjugate, conjugate transpose, and psuedo-inverse of $\vect{A}$. 
    % The real and imaginary parts of $\vect{A}$ are denoted by $\Real(\vect{A})$ and $\Imag(\vect{A})$. 
    $\vect{A}[k, \ell]$ denotes the entry of $\vect{A}$ in the $k^{\text{th}}$ row and the $\ell^{\text{th}}$ column. The same meaning is also associated with $\vect{A}_{k, \ell}$. Similarly, $\vect{A}[:, k]$ refers to the $k^\text{th}$ column of $\vect{A}$. Unspecified norm equations are $\norm{\mathbf{a}}_{2} = \mathbf{a}^* \mathbf{a}$ for vectors and the Frobenius norm $\norm{\mathbf{A}}_F = \sqrt{\text{Tr}(\mathbf{A} \vect{A}^*)}$ for matrices. We use the notation $\round{\cdot}$ for rounding to the nearest integer. We define $\mathsf{j}=\sqrt{-1}$. The operator $\mathbb{E}[\cdot]$ is used for the expectation of a random variable. Due to the notational complexity of MU-MIMO with OFDM, we will always use $u$ to refer to a specific UE, $t$ as a specific time, $k$ as a specific frequency resource, and $\nt / \nr$ to refer to a specific transmit or receive antenna.
    We use the superscript $\analog{(\cdot)}$ to refer to the analog component of hybrid variables within the MIMO system.

    The remainder of our paper is organized as follows. First, we introduce the system model and briefly highlight the SSB and CSI-RS beam management processes. We conclude this section with an overview of PMI and define the data transmission stage. Next, we introduce the proposed algorithm, $\algo$, that includes beamspace preprocessing, a multi-stage neural network, and end-to-end learning. In the final sections, we present the simulation setup and a rigorous evaluation of the various codebooks, feedback formats, and overall performance achieved by our proposed method. 
    
\section{System model}
    % Antenna/RF
    We consider one region served by a single X-MIMO base station. The BS is equipped with an $\Nta = \Nxa\times \Nya$ planar array in a hybrid architecture such that antenna elements are connected with phase shifters to radio frequency chains to form a digital dimension of size $\Nt = \Nx \times \Ny$. The phase shifters are limited to $\phaseres$ bits resolution but we assume a fully-connected structure such that every digital port is connected to each physical antenna with a separate phase shifter \cite{Alouzi2023FullHybridArray, Xue2023BMSurvey}. On the other side of the radio link, there are $U$ UEs, where $U$ is a random variable, each equipped with an $\Nr$ uniform linear array which we assume to be fully digital for this investigation. This reduces the notational complexity and allows us to focus on the BS operations, whereas UE-side optimization is left to future work. All antennas are half-wavelength spaced in the YZ planes. It remains to be seen what bands and configurations will be standardized in 6G, so we will assume channels are generated according to a model but rely on machine learning to capture channel structures rather than incorporate a specific channel model into the system. 
    % This work will also consider the effects of polarization, which is often ignored in beam management but plays an important role in beam training, especially in low and mid bands.
    The OFDM MU-MIMO channel is defined for $U$ users, over $T$ time slots, and $K$ subcarriers between each of the terminals as
        \begin{align}
            \Achan &\in \complex^{U \times T \times K \times \Nr \times \Nta} \label{eqn: Achan}.
            % \Dchan &\in \complex^{U \times T \times K \times \Nr \times \Nt}. \label{eqn: Dchan}
        \end{align}
    Note that we can generate this channel model using a clustered channel model as typically done by Sionna \cite{sionna}, 3GPP, and other raytracing or statistical models and array response vectors $ \vander{\theta, \phi}{}=\frac{1}{\sqrt{N}}[1, e^{\ji \pi \cos{\theta} \sin{\phi}}, e^{\ji 2 \pi \cos{\theta} \sin{\phi}}, ...,\ e^{\ji(N-1) \pi \cos{\theta} \sin{\phi}}]^T$ for the path direction $(\theta, \phi)$.

    With this setup in mind, we work with the following generic received signal model for signals $\vect{s}$ transmitted with digital and analog precoders $\Futk$, $\Fa$ and received with combiners $\Wutk$
    \begin{align}
        \tku{\vect{y}} &= \frac{1}{\sqrt{\Nt}} \tku{\W^*} \tku{\Achan} \Fa \sum_{\text{u}=0}^{U-1} \F_{\text{u}, t, k} \vect{s}_{\text{u}, t, k} + \tku{\W^*}\vect{N}_{t, k}. \label{eqn: rec_signal}
    \end{align}
    The noise $\vect{N}$ is modeled as independent complex Gaussian random values to account for thermal noise and the noise figure of the receiver. Our model assumes that timing and frequency synchronization have been performed, which is supported through the embedded primary synchronization and secondary synchronization sequences contained within the SSB signals. During the next subsections, we will specify how $\vect{W}, \vect{F},$ and $\vect{s}$ are determined during the SSB, CSI-RS, and data transmission stages according to an SSB codebook $\Bssb \in \complex^{}$, a CSI-RS codebook $\Bcsi \in \complex^{}$, and a predefined feedback codebook. We will assume throughout the paper that all digital beamformers are normalized according to a per-symbol power constraint, i.e. $\norm{\vect{F}}^2 =\ \Nt$ and all analog beamformers are constant modulus $|\analog{\vect{F}}_{i, j}|^2 = 1/\Nta$ for all elements $(i, j)$ to ensure comparable results across array sizes and to reflect the per-antenna power constraint that is common in practice. In addition, the analog phase shifters have a limited resolution such that each can only apply a quantized phase shift with resolution $2^{\phaseres}$. 
        % In the SSB and CSI-RS stages, $\vect{f} = \Fa\vect{f}_{t, k}$ is selected from one or more codebooks that play critical roles in beam management.
    
    We will assume a working knowledge of the beam management stages and will only briefly highlight the key equations. Readers are encouraged to review \cite{Dreifuerst2023magazine, Dreifuerst2023CodebookFeedback, Giordani3GPPBeamManagement2019} for an in-depth review of these processes in 5G. Given the identified focus on beam management for 6G \cite{Lin2023OverviewAI5G}, we will assume the same three phases (SSB, CSI-RS, and feedback) as in 5G. We also assume UEs operate similarly to sub-6GHz 5G implementations, although such simplifications are only assumed to as a simplification to focus on the BS operation. Much of the beam training and hybrid operations are not specified in the standards, so we will assume a formulation similar to classical hybrid implementations found in literature i.e. \cite{HeathOverviewSP4mmWave2016, Singh2021SurveyHybridBF, Wu2018HybridMUMIMO}.

     \begin{figure}[!t]
	    \centering
	    \includegraphics[width=3.15in]{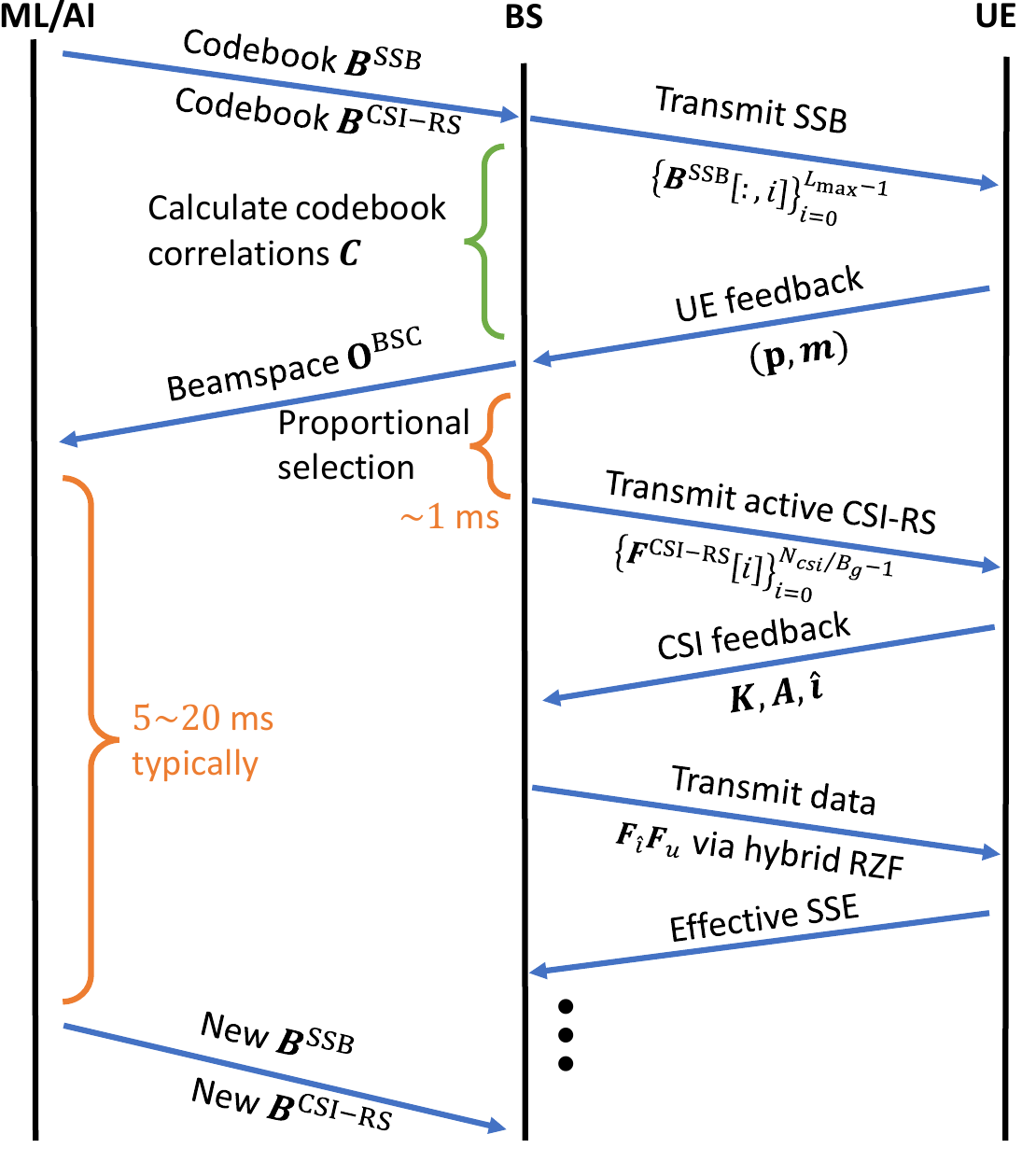}
	    \caption{A timing diagram indicating the connection between an ML/AI codebook algorithm, the BS operation, and the UE role in a downlink configuration. Deep neural networks are capable of being deployed in near real-time roles like codebook generation with time spans on the order of $5$-$20$ms. Real-time operations like precoder determination and feedback quantization are challenging due to the strict $\le 1$ms latency requirements.}
	    \label{fig: timing_diagram}
	\end{figure}
    
\subsection{Synchronization Signal Burst} \label{sec: SSB}
    The SSB process is an initial access procedure where a limited number of broadcast beams are transmitted periodically to allow UEs to connect to the network \cite{Giordani3GPPBeamManagement2019, Dreifuerst2023magazine}. Although the first SSB codebooks were designed for wide-area coverage, the concepts of ``narrow'' and ``wide'' beams do not make sense in multipath environments, especially with extremely large arrays. Instead, the SSB and CSI-RS codebooks should be designed in a hierarchical format to find more precise beamformers that improve metrics such as the received signal power or data rate. With this in mind, we overview the SSB signal model.

    The SSB signals are transmitted over a consistent band $\Kssb$ and predefined timeslots $\Tssb$ for each SSB $i$ up to the total SSB size $\Lmax$. The received signal during SSB reception before combining can be simplified from \eqref{eqn: rec_signal} to a single-stream representation of the demodulation reference signals $s^{\text{DMRS}}_{t, k} \in \complex$ as
    \begin{align}
        \vect{y}^{\text{SSB}_i}_{\ue, t, k} &=\ \frac{1}{\sqrt{K \Nt}} \Achan_{u, t, k} \f_i s^{\text{DMRS}}_{t, k} + \vect{N}_{u, t, k} \label{eqn: ssb}
    \end{align}
    where $\f_i \in \Bssb$ is the analog beamformer controlled from a single digital port. Note that a digital beamformer can also be used but it is most reasonably achieved through an analog rank-$1$ beamformer. While the limited resolution of the phase shifters $\phaseres$ in the hybrid array has a noticeable effect during multi-stream precoder design, the effects are almost imperceptible and nearly neglectable during rank-$1$ beamforming, even with low-resolution phase shifters \cite{Alouzi2023FullHybridArray}. Because the focus is on the BS operation, we will assume the UE performs digital combining to maximize the SSB reference signal received power (RSRP) as
    \begin{align}
        \RSRPiu &=\ \sum_{k\in \Kssb} \sum_{t \in \Tssb} \norm{\vect{y}^{\text{SSB}_i}_{\ue, t, k}}^2. \label{eqn: RSRP}
    \end{align}
    It is common at high frequencies to employ beam training at the UE instead of digital combining, but we leave such a scenario to future work as it is unclear whether upper-mid band 6G will operate more like sub-6GHz 5G or millimeter-wave 5G. Similarly, it is unknown if additional SSB feedback will be added in 6G, but the minimum SSB feedback corresponds to the best beam index and the RSRP of that beam. This can be aggregated at the base station as 
    \begin{align}
            \vect{p} &=\ \left\{\max_i{\RSRPiu} \right\}^{U}_{\ue=1} \label{eqn: Prsrp} \\
            \vect{m} &=\ \left\{\argmax_i{\RSRPiu} \right\}^{U}_{\ue=1}. \label{eqn: Irsrp}
    \end{align}
    The vectors $\vect{p}$ and $\vect{m}$ correspond to the best beam index \eqref{eqn: Irsrp} and the corresponding RSRP \eqref{eqn: Prsrp} for all $U$ users. After the SSB feedback, the BS can transmit CSI-RS to obtain more accurate beam alignment and channel estimates in the form of PMI.
    % We will assume digital combining by the UE to maximize the RSRP while the BS can decompose the 
    
    % and rely on simple decomposition strategies for determining the analog and digital beamforming weights \cite{Need}
    % \begin{align}
    %     \vect{f}_i &= \argmin_{\analog{\vect{F}}, \vect{F}} ||\analog{\vect{F}} \vect{F} - \Bssb_{i}||^2 \\
    %     \yssb &= \frac{1}{\sqrt{K \Nt}} \Achan_{u, t, k} \vect{f}_i
    % \end{align}
    % The challenge of codebook design is exacerbated with hybrid arrays because the beamforming vectors must be decomposed into analog and digital components. At the same time, the SSB beamforming does not need to achieve the highest RSRP. Instead, the goal is to achieve a sufficient RSRP for accurate synchronization while also providing information for the subsequent CSI-RS stage.

\subsection{CSI-RS} \label{sec: CSIRS}
    CSI-RS is the second beamforming stage where the previous SSB feedback is used to refine the beamformers to achieve a high SNR for analog beam training and channel estimation. In the current specifications (3GPP R17), the CSI-RS codebook is large, and a subset of the candidate beamformers are selected based on the SSB feedback \cite{Dreifuerst2023CodebookFeedback}. We define $\Bactive$ as the active CSI-RS codebook and the selection algorithm, $\text{Sel}(\cdot)$, that determines the $\Ncsi$ active beamformers from $\Bcsi$ based on the SSB codebook and SSBRI feedback $\vect{m}$. The subset selection can be formalized as
    \begin{align}
        \Bactive &= \text{Sel}(\vect{m}, \Bssb, \Bcsi, \Ncsi)
    \end{align}
    An important challenge during CSI-RS is that the selection operates on a short timescale--on the order of $1$ms or less. Because of this limitation, neural networks cannot easily be employed for selection; typical neural network execution times with modern hardware are on the order of $10$us to $100$ms \cite{Ren2019DLTimings, reddi2020mlperf}. Historically, the CSI-RS selection simply divides the angular space of the SSB into a proportional number of beams compared to the total allocation of CSI-RS beams, although the exact selection algorithm is left up to private implementations. Given that the learned beams do not represent simple angular spaces, we propose a similar proportional selection process using the cross-correlations of the SSB and CSI-RS codebooks. In the case of DFT beams, the proposed proportional selection produces the same results as angular division but it also extends to arbitrary, learned codebooks as well. The proportional selection algorithm begins by calculating the cross-correlations $\vect{C}$ between each of codebook entries, which can be performed while awaiting SSB feedback
    \begin{align}
        \vect{C} = \left<\Bssb_i, \Bcsi_j\right> \quad \forall i,j. \label{alg: prop_sel1}
    \end{align}
    Then the number of CSI-RS beams to be allocated to each SSB reference beam is determined as a proportion of the total number of CSI-RS based on how many users reported a given SSB beam selection $\vect{m}$
    \begin{align}
        \vect{m}^{\text{sel}} = \left\{\left\lfloor{\frac{\Ncsi}{U}\sum \vect{1}_{\vect{m}=i}} \right\rceil \right\}_{i=0}^{i=\Lmax-1}.
    \end{align}
    Finally, for each of the SSB beams, the $\vect{m}^{\text{sel}}_i$ most correlated beams are assigned to the active subset. In otherwords, the selection algorithm first determines how many of the total CSI-RS beams should ``correspond'' to each of the SSB beams, and then the most correlated CSI-RS beams for each SSB beam are selected. This can be done based on a column-wise argsort operation, the cumulative sum operation (cumsum) and vectorized selection
    \begin{align}
        \vect{A}^{\text{sorted}} &= \text{argsort}(\vect{C}) \\
        \vect{v}_i &= \text{cumsum}(\vect{m}^{\text{sel}}_{:i}) \quad \forall i \in \{0, 1, ..., \Lmax-1\}\\
        \Bactive_{\vect{v}_{i-1}:\vect{v}_i} &= \Bcsi[\vect{A}^{\text{sorted}}_{0:\vect{m}^{\text{sel}}_{i}}]. \label{alg: prop_sel2}
    \end{align}
    The result is a subset of $\Ncsi$ CSI-RS beams that are proportionally selected to be highly correlated with the active SSB beams reported.

    % \begin{algorithm}
    % \setstretch{1.3}
    % \caption{CSI-RS Proportional Selection}
    % \begin{algorithmic}[1]
    % \State Given the SSBRI $\vect{m}$, SSB codebook $\Bssb$, CSI-RS codebook $\Bcsi$ and $\Ncsi$ beams to select
    % \State $\vect{C} \leftarrow \left<\Bssb_i, \Bcsi_j\right> \quad \forall i,j$
    % \State $\vect{m}^{\text{sel}} \leftarrow \{\round{\frac{\Ncsi}{U}\sum \vect{1}_{\vect{m}==i}} \}_{i=0}^{i=\Lmax-1} \quad$ (Number of beams to allocate from each SSB)
    % \State $\Bactive \leftarrow []$
    % \State $I \leftarrow 0$
    % \For{$\ell \in \{0, 1, .. \Lmax-1 \} $}
    %     \State $\vect{a}^{\text{sorted}} \leftarrow \text{argsort}(\vect{C}_{\ell}) $
    %     \State $\Bactive_{I:I+\vect{m}^{\text{sel}}_{\ell}} \leftarrow \Bcsi[\vect{a}^{\text{sorted}}_{0:\vect{m}^{\text{sel}}_{\ell}}]$
    %     \State $I \leftarrow I + \vect{m}^{\text{sel}}_{\ell}$
    % \EndFor
    % \State Return $\Bactive$
    % \end{algorithmic} \label{alg: prop_sel}
    % \end{algorithm}

    Before discussing the received CSI-RS, it is important to understand the role of logical ports when it comes to hybrid arrays and CSI-RS. Logical ports simply correspond to a set of ports that transmit a signal over a consistent channel. It does not correspond to a specific or unique subset of analog or digital ports in a hybrid system, but rather, allows for a separation between the hardware and a representation of the channel. Although this logical-physical port separation is flexible, this formulation leaves many implementation details out of the specifications. In order to follow traditional hybrid systems, while still keeping the flexibility, we associate three parameters with the CSI-RS: $\Np$ is the total number of ports that can be configured which is typically the digital dimension of the hybrid array, $\Ncsi$ is the number of CSI-RS beamformers to be transmitted, and an additional parameter $\Bg$ which is the number of ports and beams assigned to one CSI-RS allocation. The connection between the parameters is that $\Bg$ CSI-RS beams are transmitted within one resource so that $\ceil{\Ncsi/\Bg}$ total CSI-RS resources must be allocated and up to $\floor{\Np/\Bg}$ users can be multiplexed together during the subsequent data transmission. The separation here allows for ``assigning'' portions of the logical array of size $\Bg$ to each user and users can select from $\ceil{\Ncsi/\Bg}$ resources to feedback beamformed channel estimates from. 
    
    With the active CSI-RS codebook determined, the CSI-RS process involves transmitting training symbols $\vect{s}^{\text{tr}}_{t, k}\in \complex^{\Bg \times 1}$ for channel estimation using each selected beamforming codeword, grouped into $\Bg$ beams i.e. $\F_i \subseteq \Bactive$ for each CSI-RS allocation $i$. Note that the symbols and beamformers may be one-hot or code-division multiplexed across the time-frequency resources of the CSI-RS depending on the BS-selected settings. The received signal for each UE is then
    \begin{align}
        \ycsirs_{\ue, t, k} &=\ \frac{1}{\sqrt{K \Nt}} \Achan_{u, t, k} \F_i \vect{s}^{\text{tr}}_{t, k} + \vect{N}_{u, t, k}.
    \end{align}
    From the received signal, the UE can determine the RSRP, estimated SNR, and select the CSI-RS indicator $\hat{\vect{i}}$ as
    \begin{align}
        \RSRPiu^{\text{CSI-RS}_i} &=\ \max_{\nr} \sum_{k\in \Kcsi} \sum_{t \in \Tcsi} \norm{\ycsirs_{\ue, t, k}}^2 \label{eqn: CSI_RSRP}\\
        \SNR_{i, u} &=\ \max_{\nr} \frac{1}{K \Nt} \sum_{k\in \Kcsi} \sum_{t \in \Tcsi} \frac{\norm{\widehat{\vect{H}_{u, t, k}\F_i}}^2}{\sigma^2} \label{eqn: SNR} \\
        \F_{\hat{i}_u} &= \argmax_i \SNR_{i, u} % TODO THIS IS WRONG SEE PRESENTATION
    \end{align}
    The noise power, $\sigma^2 = \mathbb{E}[\norm{\vect{N}_{u, t, k}}^2]$ can be estimated from the system parameters and zero-power CSI-RS. Compared to the SSB model \eqref{eqn: ssb}, the CSI-RS received model is multi-layer due to the need for sounding each of the $\Bg$ ports. The CSI-RS therefore must have at least $\Nr$ beams for a full-rank beamformed channel, assuming $\Nr \le \Nt$. There is a significant benefit to using groupings of a small number of beams to sound the channel as the overhead for both CSI-RS transmission and feedback quantization are reduced, and users can be multiplexed using different analog beams and well-designed digital precoders.
    
    The SNR \eqref{eqn: SNR} impacts the channel estimation efficacy and is useful for determining the initial modulation and coding scheme (MCS) for data transmission. The beamformed channel estimate is defined for a channel estimation algorithm $w(\cdot)$ as
    \begin{align}
        \widehat{\vect{H}_{u}\Fistar} = w(\ycsirshat_{u}, \vect{s}^{\text{tr}}).
    \end{align}
    There is no standardized channel estimation algorithm, but, given the limited time and power budget of mobile UEs, a least squares method is assumed herein. The resulting channel estimate at the UE is frequency selective over a set of frequency subbands $b \in \{1, ... \bwp\}$. In the final step, the channel estimates are quantized according to a feedback codebook and transmitted to the BS \cite{Dreifuerst2023magazine, Qin2023CSIFeedback}.

\subsection{Feedback} \label{sec: feedback}
    CSI feedback is a highly configurable process to achieve a careful balance of overhead and CSI accuracy. In 5G there were two categories of feedback: type-I and type-II \cite{Dreifuerst2023magazine, Qin2023CSIFeedback}. In 3GPP Release $16$ an additional format of type-II called ``enhanced type-II'' was standardized that included additional, frequency-domain compression. In general, the feedback formats are methods for quantizing the channel estimate according to a shared feedback codebook $\vect{B}^{\text{FB}}$. Type-II in particular is designed for high-resolution, MU-MIMO settings where the channel information is quantized as a set of $\Lcsi$ oversampled DFT components, with oversampling factors $\Oh\times\Ov$. We refer readers to \cite{Qin2023CSIFeedback} for a general description of the feedback formats and \cite{Dreifuerst2023CoC} for a fast implementation of type-II quantization that is important for large arrays and high resolution feedback considered in X-MIMO. 

    The BS receives a set of codebook indices $\vect{A} \in \mathbb{W}^{\bwp \times \Nr \times \Lcsi-1}$ and cophasing factors $\vect{K} \in \complex^{\bwp \times \Nr \times \Lcsi-1}$ that enable a reconstruction of the beamformed channel. Channels are reconstructed in the same way as \cite{Dreifuerst2023CodebookFeedback} i.e.
    \begin{align}
        \widehat{\vect{H}_{u, b, \nr}\Fistar} = \sum_{\ell=1}^{\Lcsi} \vect{K}_{b, \nr, \ell}\vect{B}^{\text{FB}}_{\vect{A}_{b, \nr, \ell}} \quad \forall b,\ \nr. \label{eqn: reconstruct}
    \end{align}
    The quantization error, therefore, can be reduced with large $\Lcsi$, but this introduces additional overhead. In addition, the frequency domain quantization introduces overhead that increases with a larger number of subband elements $\bwp$. The actual overhead is compressed so that the frequency-domain scaling is minimal, although the overhead of this signaling is a significant part of the precoder matrix indicator (PMI) section of the feedback report. In addition, the feedback will include information describing the rank indicator $R$ (RI), the best CSI-RS beam index $\hat{i}_u$ (CRI), and other subfields not critical to this work. The base station should then use the information from both the beam training and channel feedback to design precoders to best serve the users.

\subsection{Data transmission}
% Channel estimation and reconstruction
    In the final steps, the BS uses the CSI acquired through the UE feedback to determine the precoders to serve a set of users. The exact process is not standardized; we take a typical approach here inspired by MIMO fundamentals \cite{HeathLozano2018}.
    The analog beamforming for the hybrid array is controlled through the CRI from the feedback packet while the digital precoder is used to mitigate interference between users and streams. A typical precoder formulation is a regularized zero-forcing precoder (RZF) determined to minimize the signal-to-leakage noise ratio \cite[section 9.9]{HeathLozano2018}. The RZF precoder with perfect channel state information is
    \begin{align}
        \vect{F}_{u, t, k} =  \frac{(\sum_{i=0}^{U-1} \vect{H}_{i, t, k}^*\vect{H}_{i, t, k} + U \Nt\mathbb{E}[\vect{N}_{u, t, k}^2])^{-1} \vect{H}_{u, t, k}^*}{\norm{  \vect{H}_{i, t, k}^*\vect{H}_{i, t, k} + U \Nt\mathbb{E}[\vect{N}_{u, t, k}^2])^{-1} \vect{H}_{u, t, k}^* }}.
    \end{align}
     In the case of realistic, hybrid arrays the perfect channel state information $\vect{H}$ is replaced with the selected beamformed channels, redefined for ease of notation as $\widehat{(\vect{HF})} \equiv \{\widehat{\vect{H}_{u, b, \nr}\F_{\hat{i}_u}}\}_{u, b, \nr} \in \complex^{U \times \bwp \times \Nr \times \Bg}$. The beamformed channels arise from the beam grouping and port assignments from Section \ref{sec: CSIRS}. Without loss of generality, we assume the first ${U}_a \le \Ntd / \Bg$ users are allocated for data transmission so that the analog $\{\F_{\hat{i}_u}\}$ and digital precoders $\{\vect{F}_{u, t, k}\}$ are designed in a subarray-formulation \cite{HeathOverviewSP4mmWave2016} as
     \begin{align}
         \analog{\F} &= 
         % \begin{bmatrix}
         %    \F_{\hat{0}} & 0 & ... & 0  \\
         %    0 & \F_{\hat{1}} & 0 & ... \\
         %    0 & 0 & ... & \F_{\hat{U}} 
         % \end{bmatrix}. 
         [\F_{\hat{i}_0}, \F_{\hat{i}_1}, ..., \F_{\hat{i}_{{U}_a}} ] \label{eqn: F_RF}\\
         \F_{t, k} &= 
         \begin{bmatrix}
            \vect{F}_{0, t, k} & 0 & ... & 0  \\
            0 & \vect{F}_{1, t, k} & 0 & ... \\
            \vdots & \vdots & \ddots & \vdots\\
            0 & 0 & ... & \vect{F}_{{U}_a, t, k} 
         \end{bmatrix}. 
     \end{align}
    This formulation neatly unifies hybrid arrays, traditional precoding, and beam management techniques with the received signal model from \eqref{eqn: rec_signal}.

    On the receiver side, the UEs also perform a combining strategy based on the embedded DMRS within the signal allowing for decoding each data stream without knowing the exact precoder. In this work, we will assume an LMMSE receiver is used to maximize the signal-to-interference noise ratio (SINR) for any precoder, although the actual SINR and subsequent data rate calculations are not actually performed by the UE but are necessary to evaluate the achievable spectral efficiency in simulation. The choice of LMMSE receiver here makes sense to ensure allocated users have high performance for receiving the signal while still balancing the computational cost. The SINR at resource element ($t, k$) for a given user $u$ and data stream $r$ with equal power allocation per user is
    \begin{align}
        &\vect{I}_{u, t, k} = \bigg(\sum_{i=0}^{U-1} \vect{H}_{u, t, k} \Fistar \vect{F}_{i, t, k} (\vect{H}_{u, t, k} \Fistar \vect{F}_{i, t, k})^* + U \Nt \mathbb{E}[\vect{N}_{u, t, k}^2] \bigg)^{-1} \label{eqn: interference}\\
        &\SINR_{\ue, t, k, r} = \frac{1}{K}\frac{(\vect{H}_{u, t, k} \Fistar \vect{F}_{u, :, r})^* \vect{I}_{u, t, k} \vect{H}_{u, t, k} \Fistar \vect{F}_{u, :, r}}  
        {1 -(\vect{H}_{u, t, k} \Fistar \vect{F}_{u, :, r})^* \vect{I}_{u, t, k} \vect{H}_{u, t, k} \Fistar \vect{F}_{u, :, r}}. \label{eqn: sinr}
    \end{align}
    The SINR expression \eqref{eqn: sinr} is a ratio between the signal power fo the desired data stream $(u, r)$ compared to the noise and interference power of all other streams in \eqref{eqn: interference}. 

    We evaluate the performance in the network by the sum spectral efficiency (SSE), which is reasonable when there is not more complicated scheduling, fairness constraints, and resource allocation. The sum spectral efficiency can be multiplied by the bandwidth to get the sum rate. Assuming Gaussian signaling and treating interference as Gaussian noise, the achievable spectral efficiency, $\text{SE}_{\ue, t}$, is 
    \begin{align}
        \text{SE}_{\ue, t} = \sum_{k=0}^{K-1} \sum_{r=0}^{R-1}{\log_2(1 +\SINR_{\ue, t, k, r}}).
    \end{align}
    The SSE is a sum over the active users, but it does not account for the overhead of beam management or CSI acquisition. In the final step, we consider the effective SSE (ESSE), which accounts for the overhead due to beam training by removing the corresponding time/frequency resources due to training and feedback of the entire beam management system $(T_{\text{BM}}, K_{\text{BM}})$ from the spectral efficiency calculation
    \begin{align}
        \text{ESSE} = \sum^{U-1}_{\ue=0} \sum_{\substack{t \notin T_{\text{BM}} \\ k\notin K_{\text{BM}}}} \sum_{r=0}^{R-1} \log_2 (1 + \text{SINR}_{\ue, t, k, r}).
    \end{align}
    Although presented in a linear fashion, the SSB, CSI-RS, feedback, and data transmission typically occur over separate resources so that there are no periods of waiting in a full-buffer scenario. With the system model defined, we now present the core machine learning architecture, $\algo$ that integrates ML within the codebook design process. The proposed strategy integrates in a realistic fashion such that the system does not require additional side information and the overall architecture is designed to meet the timing constraints implicit within beam management.

% Effective SE
\section{Neural codebook design}
    \begin{figure}[!t]
	    \centering
	    \includegraphics[width=6.25in]{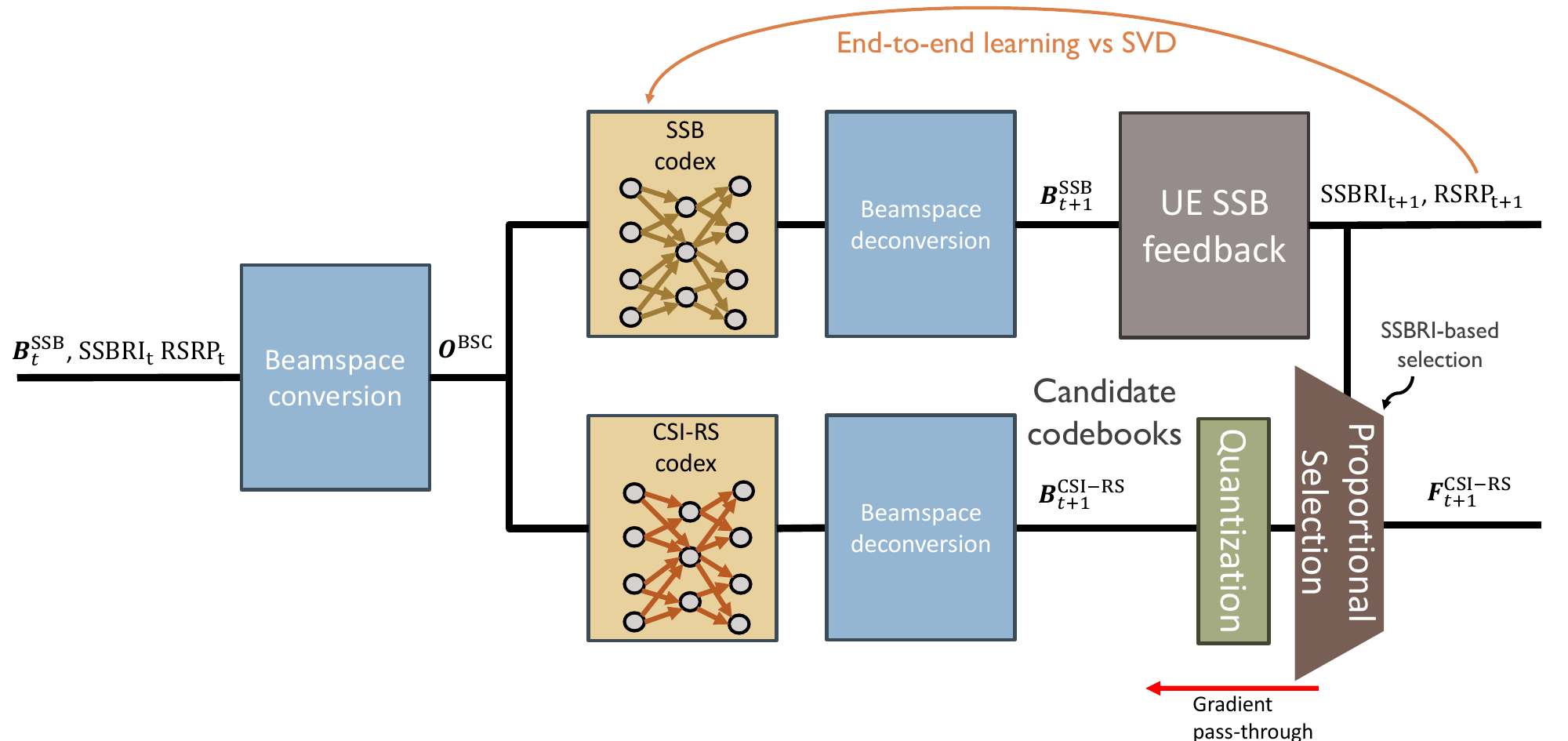}
	    \caption{A depiction of the $\algo$ algorithm and gradient propagation. The beamspace is prepared using the previous SSB codebook and feedback and fed into two parallel networks that generate the new SSB codebook and CSI-RS codebook. The CSI-RS codebook is then quantized and a proportional selection according to the new SSB feedback is used to determine the active CSI-RS codebook. Gradients are calculated according to the difference between the UE-received power and the achievable SVD received power.}
	    \label{fig: architecture}
	\end{figure}

    % We build upon our previous work with beamspace codebook design \cite{Dreifuerst2023CodebookFeedback}.
    We propose a novel approach for jointly designing the SSB and CSI-RS codebooks using AI/ML. The primary challenges with integrating machine learning into wireless domains ultimately come down to the tight timing constraints (see Figure \ref{fig: timing_diagram}) and generalizing across varying numbers of users and system configurations. Timing constraints are not particularly concerning for SSB codebook generation, as the periodicity is typically $20$ms with a minimum of $5$ms. In comparison, the CSI-RS codebook needs to be available within $1$ms following the SSB feedback, however, the CSI-RS selection depends on the SSB feedback. It is extremely challenging for modern hardware to employ deep neural networks with such tight, sequential timing constraints. Therefore, we propose the neural architecture, shown in Figure \ref{fig: architecture}, to output the SSB codebook and a set of candidate CSI-RS codebooks in parallel at each SSB period. Then we employ the proportional selection technique from \eqref{alg: prop_sel1}-\eqref{alg: prop_sel2} to determine the CSI-RS codebook from the candidate options, allowing the joint algorithm to overcome the timing limitation and still incorporate the SSB feedback. In preparation for 6G, this operation is similar to current 5G CSI-RS codebooks but extends to arbitrary and time-varying SSB and CSI-RS codebooks. The entire structure is trained in an end-to-end fashion with the goal of minimizing the power loss with respect to singular value decomposition beamforming, which is the power-maximizing strategy for a single user with perfect CSI and rank-$1$ beamforming. This ensures that we do not have to define a specific beamforming pattern train against, but rather have the network learn patterns that are close to power-maximizing. 

    The input to $\algo$ is prepared by converting the prior SSB codebook and SSB feedback to the beamspace. The input only contains previously known users, so the learning algorithm is partially regularized by new users who join the network in the upcoming SSB cycle and therefore are not included in the prior observation. The beamspace is a simple and highly extensible conversion where beamformers are converted to the angular domain with fixed resolution. This format is beneficial for translating beamformer information over time, antenna geometries, and UE numbers because the beamspace is largely unaffected by these parameters \cite{Sayeed2002Beamspace, WenDL4mMIMOCSIFeedback2018, Dreifuerst2023CodebookFeedback}.
    This ensures that the input from a set of previous beamformers represents a consistent two-dimensional grid of projections. This step was originally developed in our prior work \cite{Dreifuerst2023CodebookFeedback}.
    
    First, we define an angular transformation matrix for antenna size $N_{\text{s}}$ and $N_{\text{m}}$ angular directions as a series of array responses
    \begin{align}
        \boldsymbol{\theta}_{N_{\text{m}}} &= \frac{1}{\pi}[0, 1, ..., N_{\text{m}}-1]^T\\
        \vect{U}_{N_{\text{s}}, N_{\text{m}}} &\triangleq [\vect{a}_{N_{\text{s}}}(\boldsymbol{\theta}_0), ... \vect{a}_{N_{\text{s}}}(\boldsymbol{\theta}_{N_{\text{m}}-1})] \label{eqn: ang_matrix}.
    \end{align}
    The beamspace conversion for $N_{x0}$ azimuth directions and $N_{y0}$ elevation direction is calculated as
    \begin{align}
        \Fssb_i &= [\Bssb_{0:\Nxa, i}, ..., \Bssb_{(\Nxa-1)\Nya:\Nxa \Nya, i}] \label{eqn: reshape}\\
        \Obsc_{i} &= \vect{U}_{\Nxa, N_{x0}}^* \Fssb_i \vect{U}_{\Nya, N_{y0}} \quad \forall i.  \label{eqn: beamspace}
        % \vect{O} &= \left[\vect{F}_{\text{BSC}}, \{m_i : \sum_u \vect{m}_u=i\}, \{ \}\right]
    \end{align}
    The codebook must be reshaped from a vector of size $\Nta$ to the planar dimensions $\Nxa \times \Nya$ in \eqref{eqn: reshape} before the beamspace conversion in \eqref{eqn: beamspace} that produces the beamspace observation, $ \Obsc$. Note that the beamspace conversion is a reversible operation via the pseudo inverse so long as the number of samples in each direction is equal to or greater than the number of antennas, so we also train the network to predict the beamspace of the desired output, rather than direct beamforming coefficients to aid the learning process. The predicted beamformers are obtained within the training loop, and the computational complexity can be controlled by changing the observation sizes $N_{x0}$ and $N_{y0}$. In addition to the angular representation, the input is also concatenated with the feedback corresponding to the number of users reporting each beam and the sum RSRP. We can express this using the notation $\vect{1}_{(\cdot)}$ to refer to a vector with all $0$ entries except the indices where the subscript is true contain a $1$ and resulting in the input
    \begin{align}
        \vect{O}^{\text{BSC}} = \left\{\left[\vect{O}^{\text{BSC}}[i], \sum \vect{1}_{\vect{m}=i}, \vect{1}_{\vect{m}=i}^T\vect{p}\right]\right\}_{i=0}^{\Lmax-1}. \label{eqn: lastOBSC}
    \end{align} This input is generated during each SSB period and fed into the $\algo$ algorithm as shown in Figure \ref{fig: architecture}.

    The overall system is designed and trained in an end-to-end fashion with respect to maximizing the RSRP. End-to-end training changes the learning goal with data-driven methods to maximize or minimize a system-level metric, rather than training toward a specific output \cite{Hoydis2018E2ELearning}. In particular, we do not identify the exact beamformers we want the neural network to learn, instead relying on the network to learn the beamformers that achieve the highest RSRP. It is difficult to exactly define an ``optimal'' beamformer formulation in general for the SSB and CSI-RS steps. This arises from the fact that even with perfect CSI, the broadcast beamformers that achieve the highest SNR with a limited number of beams depend heavily on the system parameters, user channels, and number of beamformers used. By employing end-to-end learning, we instead train the neural network to learn beamformers that achieve the highest performance relative to single-user singular value decomposition (SVD) beamforming with perfect CSI. This formulation is even more advantageous in the initial access scenario where not all users are known by the system prior to codebook generation. The general learning framework is shown in Figure \ref{fig: architecture} with a loss function comparing the SVD power and the reported RSRP for each user as
    \begin{align}
        \vect{U}_u, \vect{S}_u, \vect{V}_u =&\ \vect{H}_u \quad \forall u \in \{1, ... U\} \\
        \mathcal{L}(\vect{S}, \vect{p}) =&\ \frac{1}{U} \sum_u (10\log_{10}(\vect{S}_{u, 0}^2) - 10\log_{10}(\vect{p}_u^2)) \label{eqn: loss}
    \end{align}
    where $\vect{p}$ is the RSRP from \eqref{eqn: Prsrp} or \eqref{eqn: CSI_RSRP} for the SSB or CSI-RS codebooks. Gradients are calculated from the loss function \eqref{eqn: loss} throughout the network and parameters are updated using an Adam optimizer \cite{kingma2014adam} with cyclic learning rate scheduling between $10^{-3}$ and $10^{-5}$. 

    One challenge of codebook design for X-MIMO is the massive physical array size, which results in a significant number of parameters for generating codebooks.
    % We mitigate this by making the angular transformation matrix \eqref{eqn: ang_matrix} smaller than the array size, i.e. $ N_{x0}\le \Nxa$ and $ N_{y0} \le \Nya$.
    We mitigate this with efficient neural structures like convolutional layers (See Table \ref{tb: arch}) that share parameters and efficiently operate over the beamspace representation \eqref{eqn: beamspace}. In fact, this formulation allows the neural network to learn from a universal pattern that allows for training and testing on any array geometries, so long as the beamspace dimensions are equal or greater to the antenna dimensions to prevent aliasing \cite{Dreifuerst2023CodebookFeedback}. We examine how the network generalizes across geometries in Figure \ref{fig: geometry}. It is noteworthy that further parameter reduction can be achieved using convolution transpose layers instead of a fully connected layer within the network, although we found this caused a slight performance drop in our previous work \cite{Dreifuerst2023CodebookFeedback}.

    {\rowcolors{2}{blue!10}{}
    \begin{table*}[!t]
        \caption{XBM neural architectures} \label{tb: arch}
        \centering
        \renewcommand*\arraystretch{1.15}
        % \small
        \begin{tabular}{|c|c|c|c|}
        \hline
        \textbf{Layer}       & \textbf{Primary Parameter}   & \textbf{Activation} & \textbf{Output Dimension}                \\ \hline
         % Flatten             &                              &                              & ($2\Lmax (N_{x0}+2)(N_{y0}+2)$)           \\ \hline
         Conv+Max Pool         & 128 Filters                  & ReLU                         & $\ceil{(N_{x0}+2)/2} \times \ceil{(N_{y0}+2)/2} \times 128$                                    \\ \hline
         Conv+Max Pool             & 96 Filters                     & ReLU                             & $\ceil{(N_{x0}+2)/4} \times \ceil{(N_{y0}+2)/4} \times 96$                            \\ \hline
         Dropout         & 0.3 Rate                  &                         & $\ceil{(N_{x0}+2)/4} \times \ceil{(N_{y0}+2)/4} \times 96$                            \\ \hline
         Conv             & 320 Filters                     & ReLU                             & $\ceil{(N_{x0}+2)/4} \times \ceil{(N_{y0}+2)/4} \times 320$                             \\ \hline
         Flatten         &                   &                          & $(\ceil{(N_{x0}+2)/4}) (\ceil{(N_{y0}+2)/4})(320)$                             \\ \hline
         Dropout             & 0.1 Rate                     &                              & $(\ceil{(N_{x0}+2)/4}) (\ceil{(N_{y0}+2)/4})(320)$                             \\ \hline
         Fully Conn.         & $2N_{x0}$$N_{y0}$$ N$ Neurons                 &                          & $2N_{x0} N_{y0} N$                             \\ \hline
         % Fully Conn.         & 80 Neurons                  & ReLU                         & (80)                             \\ \hline
         % Fully Conn.         & ($2N_{x0} N_{y0}\Lmax $) Neurons                 &                          & ($N_{x0}, N_{y0}, 2\Lmax $)                             \\ \hline
         Reshape             &                              &                              & $N_{x0} \times N_{y0} \times 2 N$           \\ \hline
        \end{tabular} 
        \normalsize
    \end{table*}
    }

\section{Data generation framework}
    The system-level simulation requires two stages: 1) spatially and consistent channel generation for multiple users and 2) beam management integration of the iterative steps for initial access, beam refinement, feedback, and data transmission. We start by generating raytraced channels using Sionna \cite{sionna}. In our simulations, we consider two environments (labeled A and B) with $10000$ potential users scattered over a BS sector site. The BS is equipped with an $\Nxa=16$, $\Nya=16$ planar array using a 3GPP 3D antenna model with half-wavelength spacing and $\Nx=8$ and $\Ny=4$ with $\phaseres=2$ bit phase shifters facing the serving sector region. Each user has $\Nr=4$ antennas tuned for a $10$GHz carrier frequency in environment A and tuned for a $20$GHz carrier frequency in environment B to consider low-band and high-band frequencies for upper-mid bands \cite{Holma2021XMIMO}. Users are assumed to have random mobility patterns with an average speed of $3$m/s in environment A and $10$m/s in environment B and arrays oriented vertically. Environment A models the area around the Frauenkirche in Munich with a BS placed on a building $40$m tall and Environment B models the area around the Arc de Triomphe in Paris with the BS at a height of $27$m \cite{sionna}. The user channels are sampled at $1$ms intervals over $100$MHz with $270$ resource blocks assuming $30$kHz subcarrier spacing. After a database of channels is generated, the neural network is integrated into a processing pipeline following the system model. The neural network is trained on $40,000$ training samples with a random number of users and varied locations throughout the scene. The validation dataset is drawn from the same distribution and used to determine when the network training is complete. All test results come from newly generated channels according to the test environment, which may be the same or a different environment from the training set.

    The data processing begins by first selecting a random number of active users $U \sim \mathcal{U}[4, 16]$ for the subsequent timestep and drawing that many users' channels randomly from the channel database. Then, the beamspace is calculated using the prior SSB codebook (initially set as random DFT codebook beams) with $80\%$ of the active users reporting. The remaining $20\%$ are assumed to be new users that are not known prior to the current SSB cycle and helps prevent the neural network from overfitting to known users. The new SSB and CSI-RS codebooks are generated and the current iteration begins with the SSB transmission and feedback ($\vect{p}$, $\vect{m}$) according to Section \ref{sec: SSB}. Then, the active CSI-RS codebook $\Bactive$ is selected according to the proportional correlation selection, beams are grouped into sets of length $\Pcsi$ to form beamformed channels and transmitted. During data transmission, we assume the network calculates the RZF precoders based on the feedback and estimates the expected SSE for each combination of UEs, and selects the highest SSE user set. This information is used as a simple, realistic scheduling algorithm using the CSI available. The resulting SSB RSRP, CSI-RS SNR, and ESSE metrics are used to evaluate the proposed $\algo$ from a modular and system-level context.

\section{Simulation results}
    Given the capabilities for overfitting with neural networks, we carefully evaluate the proposed system across different scenarios, array geometries, and out-of-distribution settings. From these evaluations, we seek to better understand the advantages and limitations of a neural codebook algorithm. Unless specified, the default parameters for the results are $\Lmax=16$ SSB beams, $\Ncsi=16$ CSI-RS beams, $\Bg=4$ logical ports or CSI-RS beams per CSI-RS resource, $\nRB=24$ resource blocks occupied by a CSI-RS resource, and $\bwp=8$ subbands for PMI.

    \subsection{SSB Performance}
\begin{figure}[!t]
    	    \centering
    	    \includegraphics[width=3.15in]{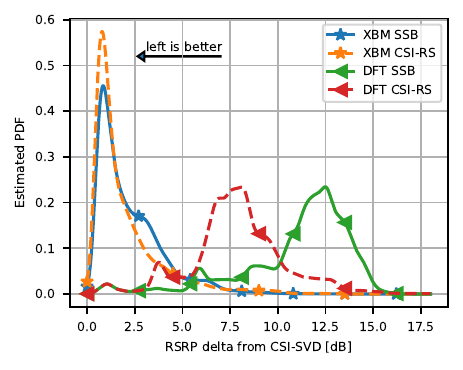}
    	    \caption{Comparison of the beamforming codebook performance relative to perfect CSI beamforming with $\Lmax=16$ and $\Ncsi=16$ beams. $5000$ sets of active users are drawn and each codebook method is used to select the SSB (solid lines) and CSI-RS (dashed lines) codebooks. The RSRP is then calculated and the difference from SVD beamforming is shown as an estimated density function. The $\algo$ codebooks are extremely effective, with more than $75\%$ of users within $3$dB of perfect CSI. Even compared to much larger DFT codebooks for CSI-RS, there is a substantial performance improvement with the proposed $\algo$ method.}
    	    \label{fig: SSB_RSRP}
    	\end{figure}

        The first evaluation focuses on the RSRP performance of the proposed algorithm. The RSRP is evaluated using both the SSB and CSI-RS codebooks, although the CSI-RS relies on the SSB feedback and should in general outperform the SSB beamforming given the additional information available to the system.
        % In particular, the SSB RSRP is typically focused on ensuring all users report an RSRP above a threshold which is often $-120$dBm \cite{DreifuerstEtalCCO}.
        In Figure \ref{fig: SSB_RSRP} we show the empirical probability distribution function (PDF) of the reported RSRP using the proposed $\algo$ as well as DFT codebooks. To compare the results, we show the relative power-loss of the RSRP achieved with the codebook method compared to SVD beamforming with perfect CSI (CSI-SVD). In the first results, we show the RSRP performance with $\Lmax=16$, which is more than supported in sub-6GHz 5G, but less than the $\Lmax=64$ supported by mmWave 5G. It can be seen that the performance of $\algo$ is almost always within $3$dB of SVD beamforming and achieves a gain of over $6$dB on average compared to DFT codebooks. Especially interesting, we see that there is very little gain from the CSI-RS with $\Ncsi=32$ for the majority of users, with only the bottom $20\%$ of users seeing noticeable improvements. 
         
\begin{figure}[!t]
    	    \centering
    	    \includegraphics[width=3.15in]{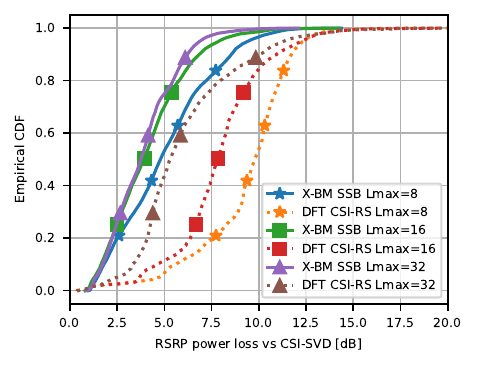}
    	    \caption{Comparison of the $\algo$ SSB RSRP with different sizes of codebooks ($\Lmax$) versus CSI-RS DFT codebooks. The DFT CSI-RS codebooks always include $\Ncsi=32$ active beams selected based on the proportional selection algorithm with the reported SSBRI. The proposed $\algo$ SSB codebooks outperform $4$ times larger SSB+CSI-RS DFT codebooks.}
    	    \label{fig: SSB_Lmax}
    	\end{figure}

        In Fig. \ref{fig: SSB_Lmax}, the performance with respect to different SSB codebook sizes $\Lmax$ is plotted. An interesting result of these simulations (Figures \ref{fig: SSB_RSRP}-\ref{fig: SSB_Lmax}) is that the $\algo$ SSB performance often outperforms even the CSI-RS performance of DFT codebooks. From a beam training perspective, there appears limited need for using CSI-RS beams when employing $\algo$ codebooks, because the SSB codebooks are already well-performing. Although a hierarchical search provides significant gains with DFT codebooks, it can be seen that the $\algo$ SSB codebook typically achieve a sufficient beam alignment. In particular, $\algo$ achieves better beam training RSRP with $\Lmax=8$ beamformers than the CSI-RS beams achieve from a $4$ times larger DFT codebook, although these results can still be improved with a larger $\algo$ codebook as shown in Figure \ref{fig: SSB_Lmax}.
    
    	% \begin{figure}[!t]
    	%     \centering
    	%     \includegraphics[width=3.15in]{figs/RSRP_delta_scene_A_256_csirs.pdf}
    	%     \caption{An evaluation of the model performance with respect to location. Obstructed locations (indoor areas not considered in the scene) are shown in black while the performance loss w.r.t. CSI-SVD is shown by the image color. We can see that only NLOS locations that are far from the BS show larger than zero error.}
    	%     \label{fig: SSB_Lmax}
    	% \end{figure}

    \subsection{CSI-RS SNR}
        Following the SSB transmission, the candidate CSI-RS codebook is selected based on the SSBRI and transmitted. The first goal is to achieve a high channel estimation SNR, while also ensuring the beamformer can be employed for hybrid beamforming with digital precoding. An inherent challenge of the design and operation of CSI-RS codebooks is the dependence on the SSB codebook and the extensive overhead associated with larger active codebooks. It is often advantageous to employ a larger number of SSB beams, which require relatively low overhead, in combination with a limited number of CSI-RS beams that consume significant wireless resources. At the same time, the SSB periodicity is higher than CSI-RS--typically about $4$ times faster--so there is additional overhead to consider. Furthermore, the design challenge is not simply to maximize SNR, but also to obtain an analog precoder that performs well for the hybrid data transmission.
        
\begin{figure}[!t]
    	    \centering
    	    \includegraphics[width=3.15in]{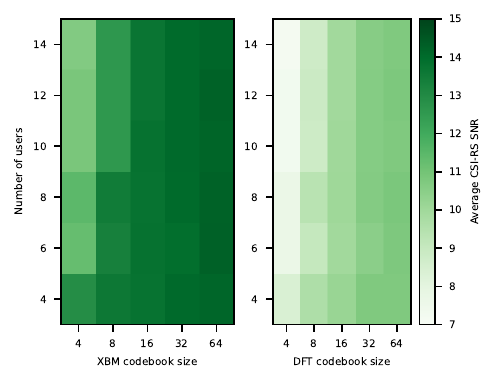}
    	    \caption{A heatmap of the average SNR during CSI-RS reception with different numbers of active CSI-RS beams $\Ncsi$ and active users $U$. While increasing the DFT codebook size provides a significant gain for more than $4$ users, the rich multipath environment enables better performance with $\algo$ codebooks, even with $1/16$ as many CSI-RS beams. This is a limitation of DFT beamformers in low- and mid-band environments.}
    	    \label{fig: CSIRS_SNR}
    	\end{figure}

        We begin by addressing the first question of achieving a high-gain CSI-RS codebook. In Figure \ref{fig: CSIRS_SNR} we compare the CSI-RS SNR obtained from classic DFT codebooks with varying sizes compared to the number of active users. We can see that when the number of users is much larger than the number of beams, the $\algo$ method performs significantly better than DFT codebooks. We can see that with just $1$ beam per active user the $\algo$ codebooks reach a consistent performance gain of about $4$dB over DFT codebooks of any size. This can be attributed to the multipath propagation environment that does not match DFT beamformers in general.
    
     \subsection{Geometry translation}
\begin{figure}[!t]
    	    \centering
    	    \includegraphics[width=3.15in]{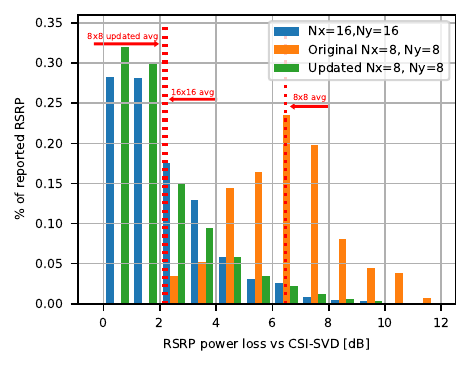}
                \caption{A comparison of two $\algo$ networks for geometry translation. The $\algo$ models are trained  using data from arrays equipped with $\Nx=\{8, 16\}$ and $\Ny = \{8, 16\}$ and beamspaces of size $N_{x0}=16, N_{y0}=16$, but both are evaluated with $\Nx=16$, $\Ny=16$ and $\Lmax=16$. Initially, there is a performance drop of about $4$dB, but fine-tuning restores the performance to that of a model trained specifically for the array geometry. This highlights the effectiveness and generalization capability of the beamspace conversion.}
    	    \label{fig: geometry}
    	\end{figure}

     % show our algorithm translating across domains very effectively
        One of the beneficial aspects of the $\algo$ structure is a natural translation across array geometries. In particular, the neural network size and parameters do not depend on the antenna array, so the same neural network can be employed or trained across geometries. Figure \ref{fig: geometry} shows a comparison of the performance of geometry translation, assuming the arrays are placed in the same scene and location. We focus on the more difficult setting of training a neural network on data from a smaller array than it is tested on. It can be seen that there is a drop in performance when testing models trained on $8\times8$ arrays in $16\times16$ settings, although the performance is still better than the DFT codebooks in Figure \ref{fig: SSB_RSRP}. In general, this is a limitation of deep learning in that does not automatically generalize well, even with carefully designed formulations.
        While the performance loss is noticeable, the benefit to the $\algo$ architecture is that the same neural network can be trained with data from different array sizes. This allows network operators to only need to collect data once and then fine-tune after selecting the hardware to maximize performance. To show the fine-tuning performance, we include the original $8\times8$ model that has been retrained on the $16\times16$ data for $30$s, highlighting how effective fine-tuning is for resolving the geometry translation loss. In this setting, fine-tuning is performed by simply training on new data with the same training steps, however, we will consider fine-tuning on imperfect CSI in Section \ref{sec: model_update}.

        % We can see that although there is a minor drop in performance compared to training directly, the overall geometry shift only causes a small impact on the unchanged model. This is a significant benefit for network operators who are considering deploying different arrays within the network without needing additional training data or retrials.
    
    \subsection{Feedback}
        The previous results have shown that ML codebooks are extremely effective for beam training, but beam alignment is not the ultimate goal of wireless networks. Effective beam training naturally ensures that a strong rank-$1$ wireless link exists, however, the actual MU-MIMO performance is less clear from the beam alignment results. In this subsection, we incorporate type-II feedback and multi-stream data transmission to determine the system-level impact of ML codebooks. We first address the general impact of the CSI-RS parameters for hybrid arrays, then look at a comparison between traditional codebooks and $\algo$ codebooks. Throughout these results, the most important metric is always the effective sum spectral efficiency, which describes the actual performance after accounting for the overhead.

\begin{figure}[!t]
    	    \centering
    	    \includegraphics[width=3.15in]{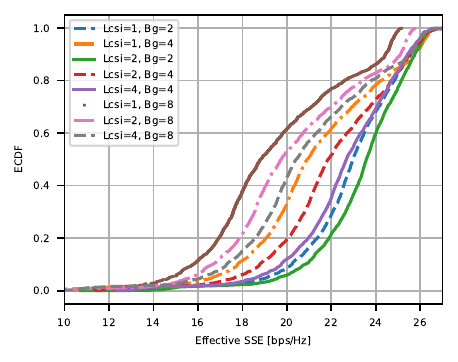}
    	    \caption{An empirical CDF chart of the ESSE performance with PMI type-II using different ratios of feedback resolution $\Lcsi$ and CSI rank $\Bg$. Each ratio is plotted with different line formats. It can be seen that there is no gain from using larger beam groupings which require more overhead, more feedback, and reduces the number of UEs that can be multiplexed.}
    	    \label{fig: PMI_typeII}
    	\end{figure}

        First, we investigate the feedback parameters using traditional DFT codebooks and hybrid arrays in Fig. \ref{fig: PMI_typeII}. These results show an interesting result: using higher rank CSI-RS requires more feedback and typically reduces the performance compared to lower rank CSI-RS. This means there is little benefit to configuring the system with many CSI-RS ports. In fact, MU-MIMO with at least one user receiving $4$ data layers is unlikely to maximize performance so even smaller beam groupings can be used. These results are specific to the simulated scenario, which includes many UEs to multiplex between and encourages scheduling more users with lower rank channels and less overhead. The total overhead depends on the number of CSI-RS resources allocated $\Ncsi/\Bg$, the beams per group $\Bg$, and the feedback and overhead associated with each CSI-RS.

\begin{figure}[!t]
    	    \centering
    	    \includegraphics[width=3.15in]{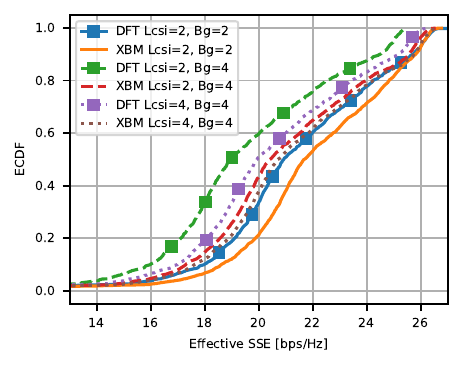}
    	    \caption{ESSE performance with CSI parameters $(\Lcsi=2, \Bg=2)$, $(\Lcsi=2, \Bg=4)$, and ($\Lcsi=4, \Bg=4)$. The $\algo$ codebooks provide a noticeable improvement over DFT codebooks, especially for the lowest $20\%$ of users. While the significant gain in beam training does result in improved ESSE, it is not expected to see gains on the order of $4$ times larger ESSE, especially with greedy scheduling.}
    	    \label{fig: CRI}
    	\end{figure}

        In the next investigation, we compare how the SSB and CSI-RS codebooks impact the overall system ESSE. The results in Figure \ref{fig: CRI} are compared again using DFT and $\algo$ codebooks with three feedback settings from the previous results, $(\Lcsi=2, \Bg=2)$, $(\Lcsi=2, \Bg=4)$, and ($\Lcsi=4, \Bg=4)$. It can be seen that the $\algo$ codebooks result in roughly a $10\%$ effective spectral efficiency improvement over DFT codebooks using the standard feedback formats. Furthermore, the gain from using limited feedback is increased with $\algo$ codebooks due to the more effective beam alignment. 
        % Based on this effective beam alignment, Figure \ref{fig: PMI_CRI_typeIII} evaluates different feedback formats including a rank-$1$ PMI-less format, standard type-II, ABC only. 
    
     %        \begin{figure}[!t]
    	%     \centering
    	%     \includegraphics[width=3.15in]{figs/ToUpdate.png}
    	%     \caption{Eff-SSE performance with 1) just SSB rank-1, 2) PMI type II, 3) just ABC, and 4) PMI type-III.}
    	%     \label{fig: PMI_CRI_typeIII}
    	% \end{figure}

    \subsection{Model updating} \label{sec: model_update}
        For the final evaluation, we evaluate how the model performs as a result of distribution shift. We consider the case that the model is initially trained in the first environment, but then the scene changes significantly causing the channel distributions to be out-of-distribution \cite{Shai2010OOD} or arising from a new context. We evaluate the model in the new environment (environment B) and then again after retraining the model using only the available CSI at the BS. In particular, we retrain on potentially erroneous CSI to characterize how effective live-model updating is using imperfect CSI. Because of the lack of perfect CSI, we employ an unsupervised learning strategy to maximize the RSRP (or minimize the negative RSRP)
        \begin{align}
            \mathcal{L}(\vect{p}) =& -\frac{1}{U} \sum_u 10\log_{10}(\vect{p}_u)^2. \label{eqn: loss_update}
        \end{align}
        This formulation is less stable as the training only seeks to maximize the RSRP without a reference point so a low RSRP is more penalizing, even if that UE cannot achieve a much higher RSRP due to the location/channel. Additionally, the gradients are calculated with respect to the quantized, estimated beamformed CSI which can include estimation and quantization error. At the same time, the training considers a very practical scenario where only the available information obtained from a live network is used for fine-tuning. 
        
\begin{figure*}[!t]
    	    \centering
    	    \subfloat[SSB \label{fig: RSRP_beams}]{\includegraphics[width=3.15in]{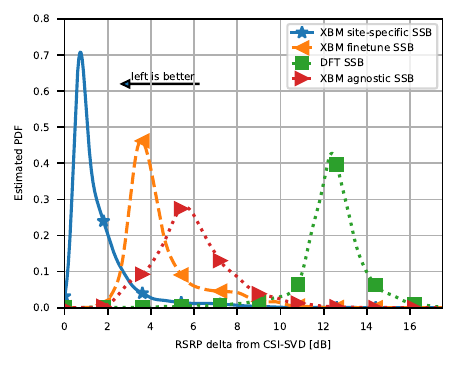}}
            \subfloat[CSI-RS \label{fig: SNR_beams}]{\includegraphics[width=3.15in]{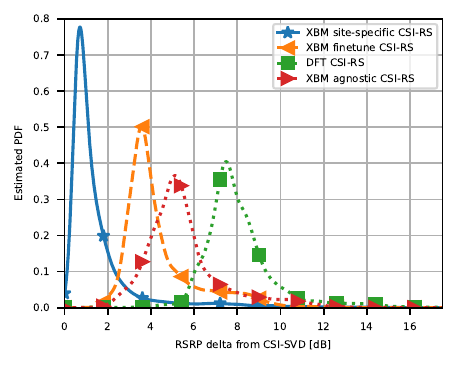}}
    	    \caption{Empirical PDF comparison of the SSB RSRP (a) and CSI-RS RSRP (b) for the $\algo$ model deployed in environment B and training on data from environment A at a different carrier frequency and user distribution. The site-specific model ($-$), fine-tuned ($--$), and site-agnostic model ($\cdot\cdot$) correspond to a model trained directly for environment B, a model trained on environment A and retrained on environment B, and the original model from environment A without retraining. Even trained on a different environment, the proposed $\algo$ method outperforms DFT codebooks and transfers to new environments efficiently.}  
         \label{fig: agnostic_scenes}
    \end{figure*}

        The results of the fine-tuning are shown in Figure \ref{fig: agnostic_scenes} which characterizes the difference between a site-specific model, the site-agnostic model, and the fine-tuned model performance between the scenarios. It can be seen that there is a significant drop in performance when deploying the initial model (site-agnostic) in the new environment, but that using the imperfect CSI and retraining with $20000$ samples for $30$s can provide an improvement, especially for the bottom $50\%$ of users. The imperfect CSI and unsupervised fine-tuning strategy, however, is not able to recover the results that could be achieved with perfect CSI shown by the site-specific results. The flexibility of deploying pre-trained models and then quickly fine-tuning is a valuable benefit for network operators, even during cyclic changes in distributions i.e. rush hour or during sporting events.
        
        % \begin{figure}[!t]
        %     \centering
        %     \includegraphics[width=3.15in]{figs/ToUpdate.png}
        %     \caption{A performance comparison of neural architectures with different beamspace sizes and architecture formulations.}
        %     \label{fig: architecure_performance}
        % \end{figure}

\section{Conclusion}
In this paper, we proposed a machine learning-assisted beam management strategy ($\algo$) for extreme MIMO. The proposed solution integrates machine learning into all stages of beam management while maintaining realistic timing and computation constraints. ML codebooks provide significant gains during beam training, especially for hybrid arrays, with an average improvement of $6$dB in received power compared to traditional methods. In fact, the learned initial access codebooks outperform two-step hierarchical search DFT codebooks in all scenarios.

The proposed $\algo$ algorithm is built upon a beamspace conversion that translates beamformers into a consistent grid representation. We integrate the beamspace conversion, quantization, transmission, and feedback into the model learning to enable an end-to-end learning framework. The proposed algorithm naturally generalizes across array geometries, user distributions, and entirely new environments. Furthermore, $\algo$ does not require changes to UE operation while increasing the achievable spectral efficiency by $10\%$. We also show that it is also capable of efficiently updating using only the partial CSI available to the network. 
% In future work, we will extend the ML-enhanced beam management concept to further include multiple base stations with limited coordination and overlapping serving regions.

Synthesizing the results of our study, we draw the following conclusions about the evolution of the upper-mid band (so-called FR3) in the next releases of 5G and into 6G. First, we find that the beam training framework can be extended to the upper midband and X-MIMO arrays. Second, DFT beams have worse performance as their narrower beamwidths (compared to sub-6GHz bands) become even more suboptimal in multipath environments. Finally, AI/ML approaches are able to design both SSB and CSI-RS codebooks that can substantially improve beam training and sum spectral efficiency within the network. Future investigations on multi-cell and interference-aware codebook design are necessary for further characterizing the benefits of AI/ML in realistic, network-level beam management.

\bibliographystyle{IEEEtran}
% {\footnotesize
\bibliography{IEEEabrv, RMD_all_refs, heath_refs_all}
% }
\end{document}